\documentstyle[12pt]{article}
\begin{document}
\begin{center}
{\large \bf LORENTZ SYMMETRY VIOLATION} \\[0pt] \vspace{0.1cm}
{\large \bf AND SUPERLUMINAL PARTICLES}\\[0pt] \vspace{0.1cm} 
{\large \bf AT FUTURE COLLIDERS}\\[0pt]
\vspace{0.5cm} {\bf Luis GONZALEZ-MESTRES}\footnote{%
E-mail: lgonzalz@vxcern.cern.ch}
\\[0pt]
\vspace{0.3cm}
{\it Laboratoire de Physique Corpusculaire, Coll\`ege de France \\
11 pl.
Marcellin-Berthelot, 75231 Paris Cedex 05, France
\\[0pt] and
\\[0pt]
Laboratoire d'Annecy-le-Vieux de Physique des Particules \\ B.P. 110 , 74941
Annecy-le-Vieux Cedex,
France} \vspace{0.8cm}
\end{center}

\begin{abstract}
If textbook Lorentz invariance is actually
a property of the equations describing a sector
of the excitations of vacuum above some critical distance scale,
several sectors of matter with different
critical speeds in vacuum can coexist and an absolute rest frame (the vacuum
rest frame)
may exist without contradicting the apparent Lorentz invariance felt by
"ordinary" particles (particles with critical speed in vacuum equal to $c$ ,
the speed of light). Sectorial Lorentz invariance, reflected by the fact that
all particles of a given dynamical sector have the same critical speed in
vacuum, will then be an expression of a fundamental sectorial symmetry
(e.g. preonic grand unification or extended supersymmetry) protecting a
parameter of the equations of motion. Furthermore, the sectorial Lorentz
symmetry may be only a low-energy limit, in the same way as the relation
$\omega $ (frequency) = $c_s$ (speed of sound) $k$ (wave vector) holds for
low-energy phonons in a crystal. In this context, phenomena
such as the absence of Greisen-Zatsepin-Kuzmin cutoff and the stability
of unstable particles at very high energy are basic properties of a wide
class of noncausal models where local Lorentz invariance is broken
introducing a fundamental length. Observable phenomena
from Lorentz symmetry violation and superluminal sectors of matter
are expected
at very short wavelength scales, even if
Lorentz symmetry violation remains invisible to standard low-energy
tests. We discuss signatures of this new physics at LHC and at other
possible future colliders.
\end{abstract}

\section{Introduction}

{\it ~~~"The impossibility to disclose experimentally the absolute motion
of the earth seems to be a general law of Nature"
\vskip 1mm
H. Poincar\'e
\vskip 3mm
"Precisely Poincar\'e proposed investigating what could be done with the
equations without altering their form. It was precisely his idea to pay
attention to the symmetry properties of the laws of Physics"
\vskip 1mm
R.P. Feynmann
\vskip 3mm
"The interpretation of geometry advocated here cannot be directly applied
to submolecular spaces... it might turn out that such an extrapolation is
just as incorrect as an extension of the concept of temperature to particles
of a solid of molecular dimensions"
\vskip 1mm
A. Einstein}
\vskip 3mm

Is relativity the result of a symmetry of the laws of Nature (Poincar\'e,
1904), therefore necessarily broken at some deeper level (Einstein, late
period), or does it reflect the existence of an absolute space-time
geometry that matter cannot escape (Einstein, early papers on relativity)?
Most textbooks teach "absolute" relativity (early Einstein papers) and
ignore the possibility of a more flexible formulation (Poincar\'e, late
Einstein thought) that we may call "relative" relativity (relativity is a
symmetry of the laws of Nature expressed by the Lorentz group: whether this
symmetry is exact or approximate must be checked experimentally at each new
energy scale). Modern dynamical systems provide many examples where Lorentz
symmetry (with a critical speed given by the properties of the system) is
a scale-dependent property which fails at the fundamental distance scale
of the system (e.g. a lattice spacing). In practical examples, the critical
speed of the apparently relativistic dynamical system is often less than  
$10^{-5}~c$ and "relativity", as felt by the dynamical system, would forbid 
particle propagation at the speed of light. Light would appear to such a 
system just like superluminal matter would appear to us. 
In this note, we would like to emphasize that high-energy
physics has reached technical ability allowing for crucial tests of the nature
of relativity combining experiments at the future machines with data from 
the highest-energy cosmic ray events.

Non-tachyonic superluminal particles ({\bf superbradyons}) 
have been discussed in previous papers 
(Gonzalez-Mestres, 1996 , 1997a and 1997b) and other papers have been
devoted to Lorentz symmetry violation (Gonzalez-Mestres, 1997c, 1997d and
1997e) as well as to its astrophysical consequences (Gonzalez-Mestres,
1997f and 1997g) and to its application to extended objects
(Gonzalez-Mestres 1997h). These papers refer also to other work in the field,
and are all submitted to the EPS-HEP97 Conference.

\section{Superluminal particles} 

Conventional tests of special relativity are performed using low-energy
phenomena. The highest momentum scale involved in nuclear magnetic 
resonance tests of special relativity is related to the energy of virtual
photons exchanged, which does not exceede the electromagnetic energy scale
$E_{em}~\approx ~\alpha _{em}~r^{-1}~\approx ~1~MeV$ , where $\alpha _{em}$
is the electromagnetic constant and $r$ the distance scale between 
two protons in a nucleus. However, the extrapolation between the 1 $MeV$ 
scale and the $1~-~100~TeV$ scale (energies to be covered by LHC and VLHC)
may involve a very large number, making compatible  
low-energy results with the possible existence of superluminal particles above 
$TeV$ scale.

Assume, for instance, that between $E~\approx ~1~MeV$ and $E~\approx ~100
~TeV$  
the mixing between an "ordinary" particle (i.e. with critical speed
in vacuum equal to the speed of light $c$ in the relativistic limit) 
of energy $E_0$ and 
a superluminal particle with mass $m_i$ , critical speed $c_i~\gg ~c$  and
energy $E_i$ is described in the vacuum rest frame
by a non-diagonal term in the energy matrix of the
form (Gonzalez-mestres, 1997c):
\equation
\epsilon ~\approx ~\epsilon _0~p~c_i~\rho~(p^2)
\endequation
where $p$ stands for momentum, 
$\epsilon _0$ is a constant describing the strength of the mixing 
and $\rho~(p^2)~=~p^2~(p^2~+~M^2~c^2)^{-1}$ accounts for a threshold effect
with $M~c^2~\approx ~100~TeV$ due to dynamics. 
Then, the correction to the energy of 
the "ordinary" particle will be $\approx ~\epsilon ^2~(E_0~-~E_i)^{-1}$
whereas the mixing angle will be $\approx ~\epsilon ~(E_0~-~E_i)^{-1}$ .
Taking the rest energy of the 
superluminal particle to be $E_{i,rest}~=~m_i~c_i^2~\approx ~1~TeV$, we get
a mixing $\approx ~0.5~\epsilon _0$ at $p~c~=~100~TeV$ , 
$\approx ~10^{-2}~\epsilon _0$ at $p~c~=~10~TeV$ and 
$\approx ~10^{-4}~\epsilon _0$
at $p~c~=~1~TeV$ . Such 
figures would clearly justify the search for superbradyons at LHC and VLHC 
($E~
\approx ~100~TeV$ per beam) machines
provided low-energy bounds do not force $~\epsilon _0$ to be too small.
With the above figures, at $p~c~=~1~MeV$ one would have a correction to 
the photon energy less than 
$\approx ~10^{-32}~\epsilon _0^2~p~c_i$ which, requiring
the correction to the photon energy not to be larger than $\approx 10^{-20}$ ,
would allow for large values of
$\epsilon _0$ if $c_i$ is less than $\approx ~10^{12}~c$ .
In any case, a wide range of values of $c_i$ and $\epsilon _0$ can be explored.

More stringent bounds may come from corrections to the quark propagator at
momenta $\approx ~100~MeV$. There, the correction to the quark energy would
be bounded only by $\approx ~10^{-24}~\epsilon _0^2~p~c_i$ and requiring it
to be less than $\approx 10^{-20}~p~c$ would be equivalent to $\epsilon _0~
<~0.1$ for $c_i~=~10^6~c$ . Obvioulsy, these estimates are rough and a detailed
calculation of nuclear parameters using the deformed relativistic kinematics 
obtained from the mixing would be required. It must be noticed that the 
situation is fundamentally different from that contemplated in the $TH\epsilon
\mu $ formalism and, in the present case, Lorentz invariance can remain 
unbroken in the low-momentum limit, as the deformation of relativistic 
kinematics for "ordinary" particles is momentum-dependent. Therefore, it may 
be a safe policy to explore all possible values of $c_i$ and $\epsilon _0$ at 
accelerators (including other possible parametrizations of $\epsilon $)
without trying to extrapolate bounds from nuclear magnetic 
resonance experiments.

The production of one or two (stable or unstable) superluminal particles 
in a high-energy accelerator experiment is potentially able to yield very
well-defined signatures through the shape of decay products
or "Cherenkov" radiation
in vacuum events (spontaneous emission of "ordinary" particles).
In the vacuum rest frame, a 
relativistic superluminal particle would have energy $E~\simeq ~p~c_i$ ,
where $c_i~\gg ~c$ is the critical speed of the particle. When decaying into
"ordinary" particles with energies $E_{\alpha }~\simeq ~p_{\alpha }~c$ 
($\alpha ~=~1,...,N)$ for a $N$-particle decay product), the initial energy
and momentum must split in such a way that very large momenta $p_{\alpha }~
\gg ~p$ are produced (in order to recover the total energy with "ordinary"
particles) but compensate to give the total momentum $p$ .This requires the
shape of the event to be exceptionally isotropic, or with two jets back 
to back, or yielding several jets with the required small total momentum.
Similar trends will arise in "Cherenkov-like" events, and remain observable 
in the laboratory frame.
It must be noticed that, if the velocity of the laboratory with respect to the
vacuum rest frame is $\approx ~10^{-3}~c$ , the laboratory velocity of
superluminal particles as measured by detectors (if ever feasible) would
be $\approx ~10^3~c$ in most cases (Gonzalez-Mestres, 1997a).
More details can be found in the papers quoted as references.

\section{Lorentz symmetry violation}

If Lorentz symmetry is broken at Planck scale or at some other 
similar fundamental length scale, it turns out (Gonzalez-Mestres, 1997d and
subsequent papers) that such a phenomenon can manifest itself at the highest 
cosmic-ray energies, typically in events above $10^{17}~eV$ . 
If Lorentz symmetry violation in the particle propagator follows a $p^2$
law beween $10^{21}~eV~c^{-1}$ and $100~MeV~c^{-1}$ , 
it can be $\approx ~1$ at the highest 
observed cosmic ray energies and $\approx ~10^{-25}$ at 
$p~c~\approx ~100~MeV$ . Furthermore, a $\approx ~1$ 
effect in particle propagators 
is not needed in order to get leading effects in data. It can be shown, for
instance, that a $\approx ~10^{-18}$ effect at $10^{17}~eV$ would be enough
to modify $\pi ^0$ lifetime above this energy, as
the term $m_{\pi }^2~c^3~(2~p)^{-1}$ ($m_{\pi }$ = pion mass)
becomes smaller than $10^{-18}~p~c$ Then, as energy increases, the particle
lifetime would become longer when measured in units of the standard 
relativistic Lorentz-dilated lifetime. It is even not impossible that
such mechanisms occur at lower energies, but this may be prevented
(Gonzalez-Mestres, 1997e) by the requirement that cosmic rays with energies
below $\approx ~3.10^{20}~eV$ lose most of their energy in the atmosphere.
Detailed calculations adapted to each specific model should be performed. 

A particularly appealing scenario, in view of experimental tests, would be 
the one where Lorentz symmetry violation has full strength at Planck scale.
Then, in most nonlocal models we expect (Gonzalez-Mestres, 1997d and 1997e) 
that the expression for energy in terms of momentum, in the limit
$k~a~\ll~1$ ($k$ = wave vector, $a$ = fundamental distance
scale), contain an extra 
term $\Delta ~E~\simeq ~-~\alpha ~(k~a)^2~p$ 
where $\alpha ~\approx 0.1~-~0.01$~.
Writing $\Delta ~E~\simeq ~\Gamma ~(k)~p$~, we have 
$\Gamma ~(k)~\simeq ~\Gamma _0~k^2$ where $\Gamma _0~=~-~\alpha ~a^2$~.
Such a behaviour is not incompatible with a possible gravitational origin of
the deformation, although gravity would not necessarily be a fundamental force 
at Planck scale and the graviton may be a composite object
made (like all other gauge bosons and presently known matter fields) of
more fundamental (perhaps superluminal) matter interacting at scales smaller
than $a$ (e.g. Gonzalez-Mestres, 1997d). String-like models using Planck scale
may then describe composite objects made of superluminal matter.

Again, very
high-energy accelerator experiments 
(especially with protons and nuclei) can play a crucial role. Contrary to the
previous case, they should now be performed in the very-forward region. At
LHC, FELIX could provide a crucial check of special relativity by comparing
its data with cosmic-ray data in the $\approx ~10^{16}~-~10^{17}~eV$ region.
VLHC experiments would be expected to lead to fundamental studies in the
kinematical region which, according to special relativity, would be equivalent
to the collisions of $\approx ~10^{19}~eV$ cosmic protons. With a 700 $TeV$
per beam $p~-~p$ machine, it would be possible to compare the very-forward 
region of collisions with those of cosmic protons at energies up to 
$\approx 10^{21}~eV$~. Thus, it seems necessary that all very high-energy
collider programs allow for an experiment able to cover secondary particles
in the far-forward and far-backward regions. A model-independent way to 
test Lorentz symmetry between collider and cosmic-ray data sould be carefully
elaborated, but the basic phenomena involved in the case of Lorentz symmetry
violation can be (Gonzalez-Mestres, 1997d and 1997h):

{\it i)} failure of the standard parton model (in any version, even 
incorporating radiative corrections);

{\it ii)} failure of the relativistic formulae for Lorentz contraction
and time dilation;

{\it iii)} longer than predicted lifetimes for some of the produced particles
(e.g. the $\pi ^0$).

The role of high-precision data from accelerators would then be crucial to
establish the existence of such phenomena in the equivalent cosmic-ray
events. In this way, it would in particular be possible to perform unique 
tests of special relativity involving possible violations coming from 
phenomena at some fundamental scale close to Planck scale, and even to 
determine the basic parameters of Lorentz symmetry violation (e.g. of 
deformed kinematics). 

\vskip 6mm
{\bf References}
\vskip 4mm
\noindent
Gonzalez-Mestres, L., "Physical and Cosmological Implications of a Possible
Class of Particles Able to Travel Faster than Light", contribution to the
28$^{th}$ International Conference on High-Energy Physics, Warsaw July 1996 .
Paper hep-ph/9610474 of LANL (Los Alamos) electronic archive (1996).\newline
Gonzalez-Mestres, L., "Space, Time and Superluminal Particles",
paper physics/9702026 of LANL electronic archive (1997a).\newline
Gonzalez-Mestres, L., "Superluminal Particles and High-Energy Cosmic Rays",
paper physics/9705032 of LANL electronic archive (1997b).\newline 
Gonzalez-Mestres, L., "Lorentz Invariance and Superluminal Particles",
paper physics/9703020 of LANL electronic archive (1997c).\newline
Gonzalez-Mestres, L., "Vacuum Structure, Lorentz Symmetry and Superluminal
Particles", paper physics/9704017 of LANL
electronic archive (1997d).\newline
Gonzalez-Mestres, L., "Lorentz Symmetry Violation and Very High-Energy
Cross Sections", paper physics/9706022 of LANL electronic archive (1997e).
\newline
Gonzalez-Mestres, L., "Absence of Greisen-Zatsepin-Kuzmin Cutoff
and Stability of Unstable Particles at Very High Energy,
as a Consequence of Lorentz Symmetry Violation", Proceedings of ICRC97 ,
Vol. 6 , p. 113 . Paper
physics/9705031 of LANL electronic archive (1997f).\newline
Gonzalez-Mestres, L., "Possible Effects of Lorentz Symmetry Violation on the
Interaction Properties of Very High-Energy Cosmic Rays", paper
physics/9706032 of LANL electronic archive (1997g).\newline
Gonzalez-Mestres, L., "High-Energy Nuclear Physics with Lorentz Symmetry
Violation", paper nucl-th/9708028 of LANL electronic archive (1997h).\newline
\end{document}